# Significance of Natural Scene Statistics in Understanding the Anisotropies of Perceptual Filling-in at the Blind Spot

Rajani Raman* and Sandip Sarkar
Saha Institute of Nuclear Physics, HBNI, 1/AF, Bidhannagar, Kolkata-700064, India

* rajani.raman@saha.ac.in

## Abstract

Psychophysical experiments reveal our horizontal preference in perceptual filling-in at the blind spot. On the other hand, vertical preference is exhibited in the case of tolerance in filling-in. What causes this anisotropy in our perception? Building upon the general notion, that the functional properties of the early visual system are shaped by the innate specification as well as the statistics of the environment, we reasoned that the anisotropy in filling-in could be understood in terms of anisotropy in orientation distribution inherent to natural scene statistics. We examined this proposition by investigating filling-in of bar stimuli on a Hierarchical Predictive Coding model network. In response to bar stimuli, the model network, trained with natural images, exhibited anisotropic filling-in performance at the blind spot similar to reported in psychophysical experiments i.e. horizontal preference in filling-in and vertical preference in tolerance of filling-in. We suggest that the over-representation of horizontal contours in the natural scene contribute to the observed horizontal superiority while the broader distribution of vertical contours contributes to the observed vertical superiority in tolerance. These results indicate that natural scene statistics plays a significant role in determining the filling-in performance at the blind spot and shaping the associated anisotropies.

## Introduction

When two aligned bars are presented on opposite sides of the blind spot such that the gap fully falls inside the blind spot, the bars are usually perceived as a continuous one. Even though we do not receive any signal related to the bar from the blind spot region, our brain by some means fills-in the information resulting in a perception of a long continuous bar [1,2]. This phenomenon is generally referred to as perceptual completion or filling-in. Psychophysical investigations have revealed that the nature of bar filling-in depends on various stimulus attributes (e.g. length, alignment and orientation difference).

In studies related to filling-in at the blind spot[3], it has demonstrated that a certain minimum length (of bar-pair extended beyond the blind spot) is required in order for filling-in to occur. Moreover, this minimum length depends upon the orientation configuration of the bar-pair (horizontal/vertical); and additionally, for the horizontal configuration, relatively shorter length is required for filling-in to occur. It was also revealed that for the identical length, horizontal configuration tends to have better filling-in over vertical configuration. This phenomenon is designated as anisotropy in filling-in.

In his other studies[4,5], the presence of anisotropy was observed in the tolerance of filling-in. However, contrary to the conventional horizontal dominance in filling-in, in this case, vertical dominance was observed; vertical bar–pair exhibited greater tolerance to the difference in alignment or orientations for perceptual filling-in to occur. This phenomenon is designated as anisotropy in tolerance of filling-in.

These psychophysical investigations[3,4] suggest that the perceptual filling-in depends upon stimulus orientation configuration along with the stimulus attributes.

Other than blind spot filling-in, anisotropy has also been reported in other visual phenomena related to orientation perception. Studies with grating stimuli show that visual system is biased toward cardinal (horizontal and vertical) orientation compared to oblique[6]. This effect is known as 'oblique effect'. On the other hand, studies involving natural broadband stimuli reveal the opposite where oblique orientations have upper hand over cardinal ones[7–9]. This phenomenon is known as 'horizontal effect'. These studies brought out the differences in bias between horizontal and vertical orientation and demonstrated that our visual system favors horizontal configuration over vertical. It has been suggested [6,8,10–12] that the statistics of natural scenes is primarily responsible for the emergence of anisotropy in the orientation perception. This has been supported by image analysis[8,10,13–15], which revealed that the orientation content in natural scenes is biased more towards horizontal than vertical; and the least bias is observed towards the oblique. This asymmetry raises a logical question whether the orientation selective neurons in the cortex are influenced by the prevalence of horizontal orientation in the environment. Indeed, it has been demonstrated experimentally[16–19] that adult ferret and cat V1 contains an over-representation of neurons coding horizontal orientations. Studies using fMRI also show[11,12] the anisotropic preference of the human visual cortex to orientation selectivity.

The knowledge of environmental statistics and the related over-representation of neurons in V1 throw some light on the possible causes of anisotropies observed in the perceptual judgments in orientation perception. However, it does not address the phenomenon of perceptual of filling-in at the blind spot and the associated anisotropy related to filling-in. In a different context[4,5], the role of vernier acuity and elliptical shape of receptive fields of neurons was speculated in the anisotropy of perpetual filling-in. Moreover, these studies also speculate different processes for anisotropy observed in different types of tolerances of filling-in. However, this speculation neither explains different processes nor fits with a general computational mechanism of visual processing.

Very recently[20] it has been shown that bar (shifting and misaligned bar) filling-in phenomena at the blind spot could be explained by considering the inherent prediction correction mechanism of Hierarchical Predictive Coding (HPC)[21] (as the computational principle of the cortex). It was argued that in the absence of any feedforward information (due to the absence of sensory input corresponding to the blind spot region), top-down prediction dominates the filling-in of the discontinuity. The nature of filling-in, on the other hand, was in accordance with the learned internal model. For proper prediction of the input bar stimuli, it was necessary for the top-down mechanism to predict two separate bars (aligned or misaligned). Instead, in both the cases, top-down mechanism favored the presence of a single continuous bar (resulting in the filling-in at the blind spot), which was the dominant feature of the internal model learned via training with natural images.

These results suggest two very important aspects of filling-in. Firstly, filling-in at the blind spot is the outcome of the prediction-correction mechanism of the cortex and secondly, the nature of filling-in is guided by the abundance of objects present in the natural scene. However, these findings cannot explain the orientation specific anisotropies in filling-in at the blind spot, where human observers reported horizontal superiority in filling-in and vertical superiority in the tolerance of filling-in. Moreover, studies with cortical neurons demonstrated its orientation selective stability in response to selective perturbation induced by adaptation[13]. This is attributed to the anisotropic distribution of local inputs to the orientation selective neurons i.e., a narrower distribution of local inputs to the neurons makes it more stable compared to the neurons having a broader distribution of local inputs. However, the significance of these findings concerning anisotropy is not known in the context of filling-in.

We reasoned that the inherent anisotropy of natural scene could be responsible for the emergence of anisotropy in perceptual filling-in including anisotropy in tolerance of filling-in. We hypothesized that the over-representation of orientation preference in the natural scene contributes to the observed anisotropy in filling-in and the nature of orientation preference distribution determines the observed anisotropy in tolerance of filling-in.

To test these propositions, we have investigated three cases of bar filling-in at the blind spot via simulation studies in a model network[24] in the light of Hierarchical Predictive Coding scheme. We used expanding, misaligned, and rotating bar as the input stimuli in horizontal and vertical configuration. In response to these input stimuli, the model network exhibited anisotropy in filling-in as well as anisotropy in tolerance of filling-in, which corroborate the findings of psychophysical experiments with human observers.

## Results

The objective of this study is to test the hypothesis that the prevalence of certain features in natural scenes is capable of providing a mechanistic explanation of anisotropy related to the perceptual filling-in reported by human observers. Our objective is summarised in Fig. 1, where we have schematically depicted the proposition that there is a link between the anisotropy present in the natural scene and the anisotropy reported in perceptual filling-in investigations. This supports the general speculation[6,8,10–12] that orientation anisotropy in natural scene plays a significant role in determining the anisotropy in the cortex as well as the anisotropy in perceptual orientation preference. As a premise, we first explored the capability of the HPC model network to learn the anisotropic distribution of features present in the natural image via training, which will validate the previously known results. Then we went on to investigate whether the learned statistics (learned internal model) could explain the anisotropy in filling-in and the anisotropy in tolerance of filling-in reported in other psychophysical studies.

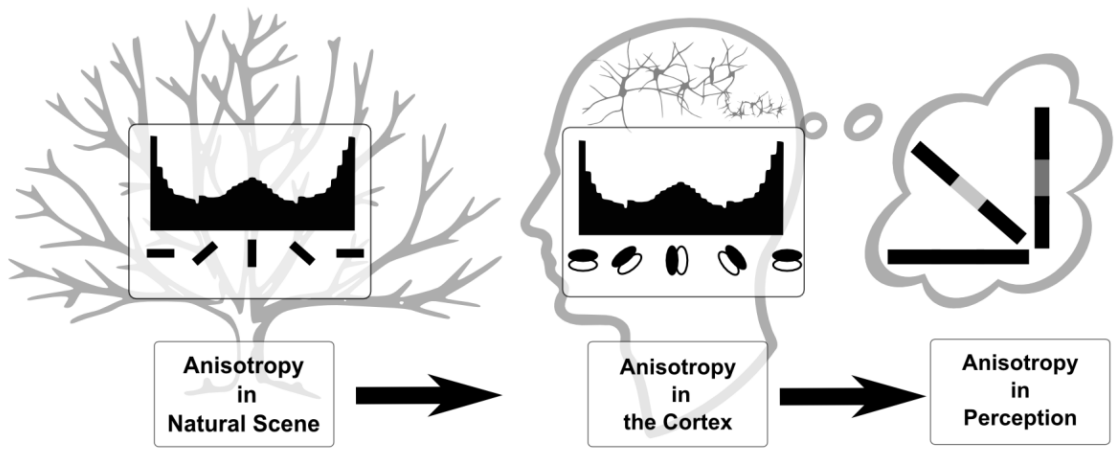

*Figure 1. Anisotropy in the natural scene, the cortex, and perception:* Our aim is schematically presented in this diagram. We want to establish the link between the anisotropy in the contours in the natural scene, orientation preference of neurons in the cortex and orientation bias in human perception.

The HPC model network considered in this study is similar to the one described in a recent investigation[20] (details are given in the method section). The network was trained with hundreds of thousands of natural image patches in one cycle. To perform the investigations with statistical rigors, we performed 40 cycles of training. As reported in several studies[20–22], each training set yielded the Gabor-like weighting profiles at level 1 (Fig. 2a) distributed in a different orientation and spatial frequency, which resembles the simple cell receptive field at V1. Level 2 weighting profiles resemble more abstract features (corner, curves, long bar etc.) as reported in recent studies[20].

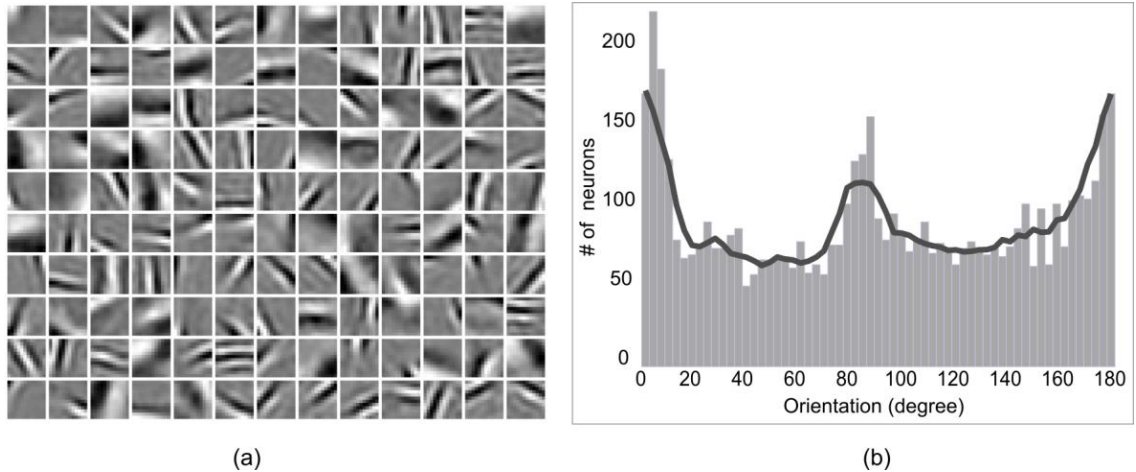

*Figure 2. Anisotropy in orientation selectivity* (a) learned weighting profiles of 130 neurons at one of the 9 modules at level 1 after a single training. (b) Orientation distribution at level 1 for all the neurons (130 X 9). The envelope (continuous line) is obtained from the average of 7 bins of the histogram.

To investigate the presence of any anisotropy, we measured the orientation tuning distribution of the trained neurons in V1. To do these, we utilised bar stimulus of different orientation and frequency and determined the orientation tuning of a particular neuron by registering their optimal response. Fig. 2b shows the distribution of orientation tuning of neurons in V1. It is evident from the distribution that larger number of neurons are oriented towards the horizontal orientation, followed by vertical and then non-cardinal orientation. This anisotropic distribution is very much in-line with the reported anisotropy of orientation distribution in natural scenes[10,13,23] and orientation tuning distribution of neurons in primary visual cortex[11,12,17,19,24].

**Anisotropy in filling-in**

To investigate the anisotropy in filling-in, the learned network was exposed to a pair of expanding bar segments, placed as shown in Fig. 3a, oriented in the horizontal direction. One end of both bars was fixed and other ends were free to expand together in sync as described in the Fig. 3. The network was also stimulated with stimuli oriented in the vertical direction (not shown). The responses of PE neurons were recorded as a function of bar extension (length) for both orientation configurations. This process was repeated 40 times with 40 different training cycles. Investigations with different training can be considered analogous to the psychophysical investigation performed on different participants (human), which leads to more statistical rigors in results. All the subsequent investigations reported in this study follow the same number of repetitions. From these simulated responses, equivalent "perceptual images" were reconstructed, which are shown in Fig. 4a for both horizontal (top row) and vertical configurations (bottom row).

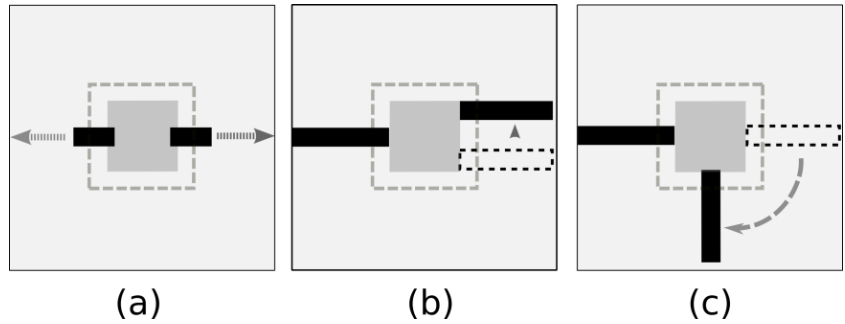

*Figure 3. Stimuli,* (a) Expanding bar stimulus: Two bar stimuli are shown at the opposite end of the blind spot, which is indicated with the gray square ($8 \times 8$ pixels) in the center. The dotted square ($12 \times 12$ pixels) denotes the area exposed to the central module (called BS module) of one of the nine level 1 modules (see Methods). One end of both bars was fixed inside the blind spot, whereas other ends were expanding together in sync in steps of one pixel in opposite directions. Extension of bars has been measured from the border of blind spot. (b) Misaligned bar stimulus: The bar at the left side of the blind spot remains fixed while the right side bar moves in the vertical direction in steps of one pixel every time (c) Rotating bar stimulus: In this case, the left side bar remains fixed but the bar at right side rotates in steps of 10 degrees.

To quantify the filling-in, pixel values in the middle (central $2 \times 2$ pixel wide region in the blind spot, indicated in the small red square in Fig. 4a) of the perceptual image were averaged. We define this average as the 'filling-in-value', where more negative 'filling-in-value' indicates better filling in. We obtained this response values from all the perceptual images corresponding to 40 training for the given factors (bar extension and configuration).

Fig. 4b shows the plot of 'filling-in-value' as a function of the bar extension for both configurations (horizontal and vertical). Inspection of Fig. 4b shows that the filling-in starts improving when the lengths of the bar segments exceed a certain minimum. This can be visualised from the perceptual images (Fig. 4a) where beyond a certain minimum length, the bars appear continuous. This result exhibits the 'minimum-length requirement'[3] properties of filling-in. The comparative plots of filling-in-value for horizontal and vertical configuration in Fig.4b, shows that for a particular filling-in value the extension of the horizontal bar remains shorter which indicates that the minimum critical length needed for the onset of filling-in would be lesser for the horizontal configuration. Moreover, for the equal bar extension, the filling-in performance is better (more negative 'filling-in-value') for the horizontal case. This anisotropic property is in agreement with psychophysical studies[3].

To validate our results, a two-way ANOVA was conducted that examined the significance of effect of degree of bar extension and the configuration (horizontal /vertical) on the filling-in-values. We found that the effect of extension [F (10,858) =933.93, p=0)], configuration [F (1,858) =585, p=0)], and, the interaction between them [F (10,858) =24.09, p=0)] was significant.

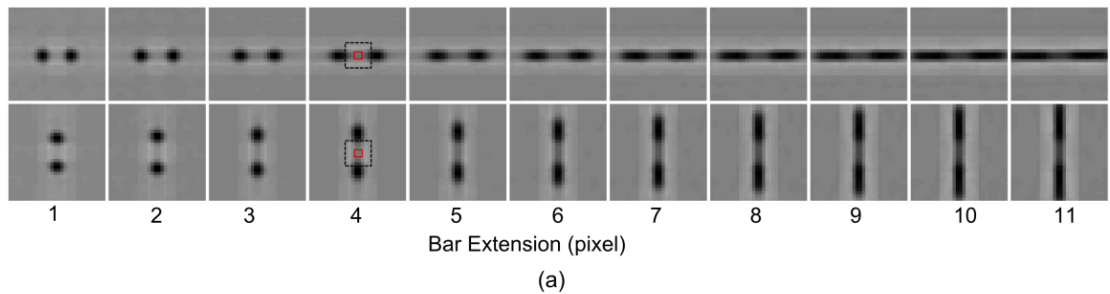

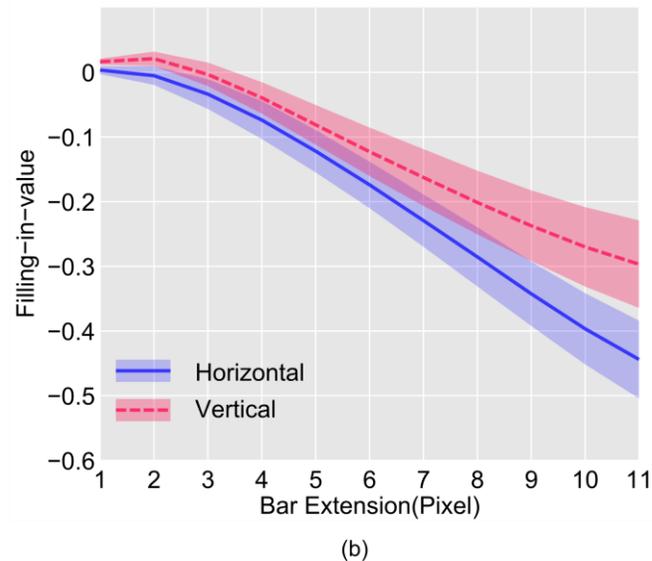

*Figure 4. Filling-in anisotropy.* (a) Perceptually equivalent images are shown, which are generated from the response of PE neurons while the network was stimulated with stimuli depicted in Fig. 3a. the dotted black square indicates the blind spot extension while small red square indicates the area ($2 \times 2$ pixel) from where filling-in-value was obtained (b) Plot of 'filling-in-value' in BS area of the images in (a) as a function of bar extension measured from the edge of the blind spot. The lines represent the average and the shaded portion indicates the standard deviation for the 40 training set.

## Anisotropy in filling-in tolerance

### Anisotropy in misalignment tolerance

For this study, the model network was exposed to a pair of bar segments placed on both sides of the blind spot and this is repeated separately for horizontal and vertical configuration. The arrangement for the horizontal case is shown in Fig. 3b. One bar was kept fixed at one side of the blind spot while the position of the other one was shifted vertically in small steps to vary the misalignments. The response of PE neurons in BS module was recorded with changing misalignment and the perceptually equivalent images were generated from these responses, which are shown in Fig. 5a (top row). Likewise, the vertical configuration gave rise the perceptual equivalent images shown in Fig. 5b (bottom row)

The images show that, in both configurations, the filling-in is best in the case of perfect alignment but deteriorates with increasing misalignment. Inspection of 'filling-in-value' plotted in Fig. 5b show that it is more negative (better filling-in) for the horizontal configuration compared to that of the vertical one, which is the signature of anisotropy of filling-in, as we have already discussed in the previous section.

However, we can also observe that the slope of the curves is higher for the horizontal case. This indicates that the rate of change of the 'filling-in-value', for the horizontal orientation, is more sensitive to the change in misalignment. In other words, filling-in, in the case of vertical orientation, is more tolerant to misalignment compared to that of the horizontal orientation. This result could be considered as a signature of anisotropy of filling-in tolerance.

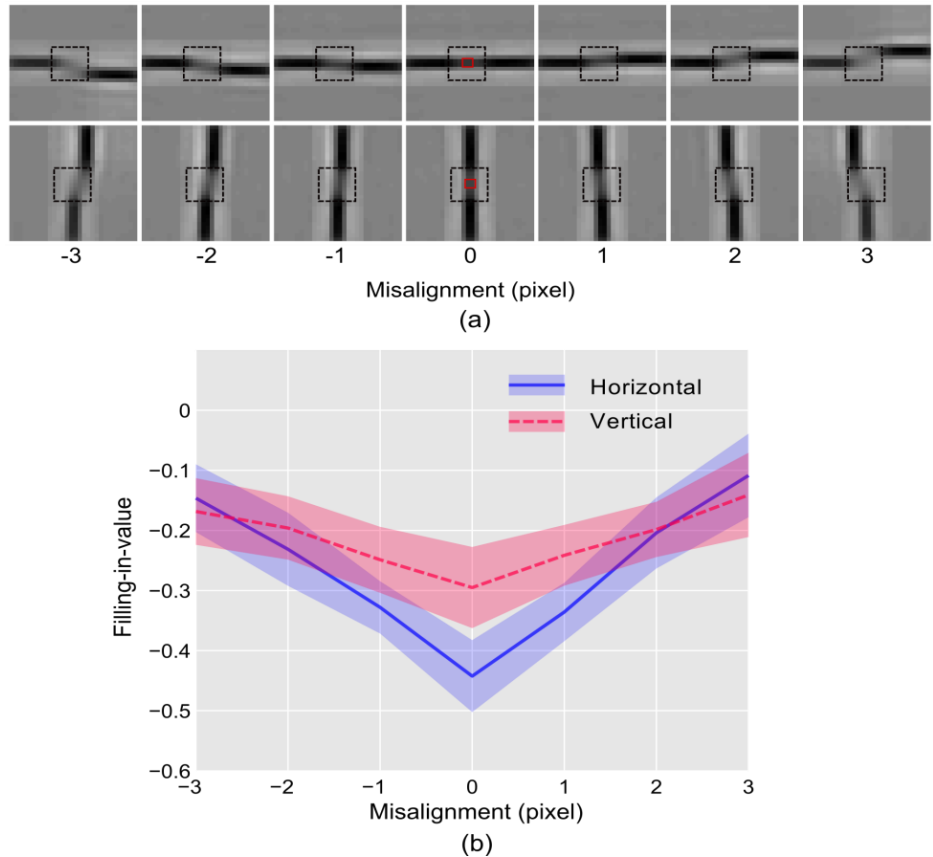

*Figure 5. Anisotropy in tolerance of filing-in of misaligned bars. (a) Perceptually equivalent images are shown, which are generated from the response of PE neurons while the network was stimulated with stimuli depicted in Fig. 3b. (b) The plot of 'filling-in-value' in BS area of the images in (a) as a function of misalignment between the bars. Convention for lines and the shades are as described in Fig. 4b.*

A two-way ANOVA was conducted that examined the significance of effect of degree of bar misalignment and the configuration on the filling-in-values. We found that the effect of misalignment [$F(6,546) = 175.91$, $p < 0.001$], configuration [$F(1,546) = 81.96$, $p < 0.001$], and, the interaction between them [$F(6,546) = 26.53$, $p < 0.001$] was significant.

**Anisotropy in disorientation tolerance**

The focus of this study was to investigate the anisotropy of tolerance of filling-in for orientation difference of two bar segments placed on both sides of the blind spot in horizontal and vertical configuration. The configuration for the horizontal case is shown in Fig 2c. The stimulus consists of a fixed bar and a rotating bar. The fixed bar is placed horizontally for the horizontal configuration and vertically for the vertical configuration. The other bar, the test bar, was rotated in steps of 10 degrees from the aligned position (0-degree difference in orientation) to the perpendicular position (90 degrees difference in orientation). The perceptual images, generated from the recordings of PE neurons, are shown in Fig. 6a for both horizontal (top row) and vertical cases (bottom row).

As expected for the both configurations, the filling-in performance was better for the aligned bars but it deteriorated with increasing difference in orientation (Fig. 6b). It is also evident that the 'filling-in-value' is more negative (indicating better filling-in), in horizontal case, throughout the entire range of difference (in orientation) from 0 degrees to 60 degrees and thereafter, the difference becomes indistinguishable. The results show that the horizontal configuration favors filling-in but exhibit more sensitivity to the changes in orientation difference (less tolerant); on the other hand, the vertical configuration is little less favorable for filling-in but is less sensitive to the changes in orientation difference (more tolerant).

A two-way ANOVA was conducted that examined the significance of the effect of degree of bar disorientation and the configuration on the filling-in-values. We found that the effect of disorientation [$F(9,780) = 334.4$, $p < 0.001$)], configuration [$F(1,780) = 104.66$, $p < 0.001$)], and, the interaction between them [$F(9,780) = 13.12$, $p < 0.001$)] was significant.

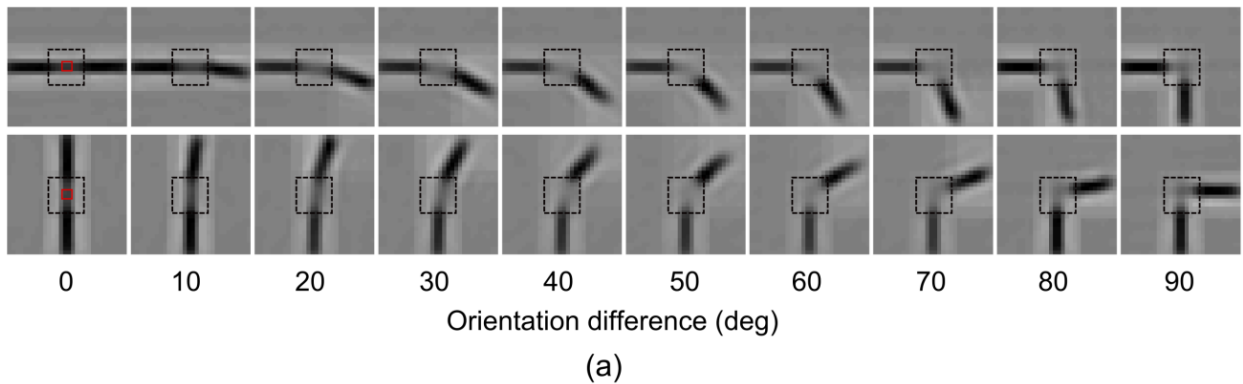

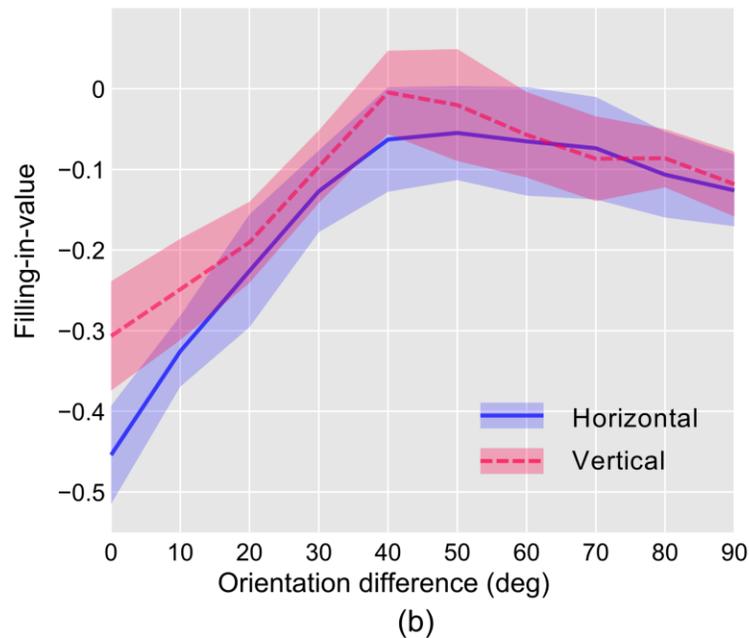

*Figure 6. Anisotropy in tolerance of filing-in of disoriented bars.* (a) Perceptually equivalent images are shown, which are generated from the response of PE neurons while the network was stimulated with stimuli depicted in

*Fig. 3c. (b) The plot of 'filling-in-value' in BS area of the images in (a) as a function of orientation difference between the bars. Convention for lines and the shades are as described in Fig. 4b.*

**Comparison with the psychophysical results**

In this study, we have conceived a general notion of the tolerance of filling-in as a rate of change of filling-in-value, with increasing difference in attributes. Faster change (higher rate) indicates lesser tolerance. This is advantageous because one can predict the tolerance by inspecting the slope of the curve representing the changing filling-in-value, which is available from the simulation study. Psychophysical studies[4,5], on the other hand, have defined tolerance of the filling-in as a maximum difference in attribute above which filling-in is not perceivable. This is completely compatible with the outcome of psychophysical experiments where the participants were asked to judge whether the line segments perceived as continuous or discontinuous.

It can be shown that our results also corroborate the results reported in[4,5]. To do this we need to normalise our results as shown in Fig. 7. This operation results in the plots shown in Fig. 7. To compare the results, we have introduced a limit at -0.65 to represent an artificial threshold above which filling-in does not happen. In line with the definition of tolerance (maximum difference for which filling-in cannot be perceived) compatible with psychophysical experiments, tolerances are represented by horizontal bars drawn at the bottom of the Fig. 7, where the length of bars gives the tolerance. This clearly shows the vertical dominance in the case of tolerance of filling-in in both the cases. Moreover, we can also observe that the relative difference (horizontal vs vertical) in tolerance is larger for misalignment in comparison to that of orientation difference. The qualitative nature of these results is in agreement with the results shown in[4,5].

The magnitude of the threshold was taken from the measurements of neural responses during bar filling-in reported in[2,25]. It is shown that the average response of neurons (BS eye) varies from 25% to 75% (in comparison to that of the neurons in the other eye) when filling-in occurs. Therefore, we have set the threshold at 50%, which after normalization becomes approximately 65%.

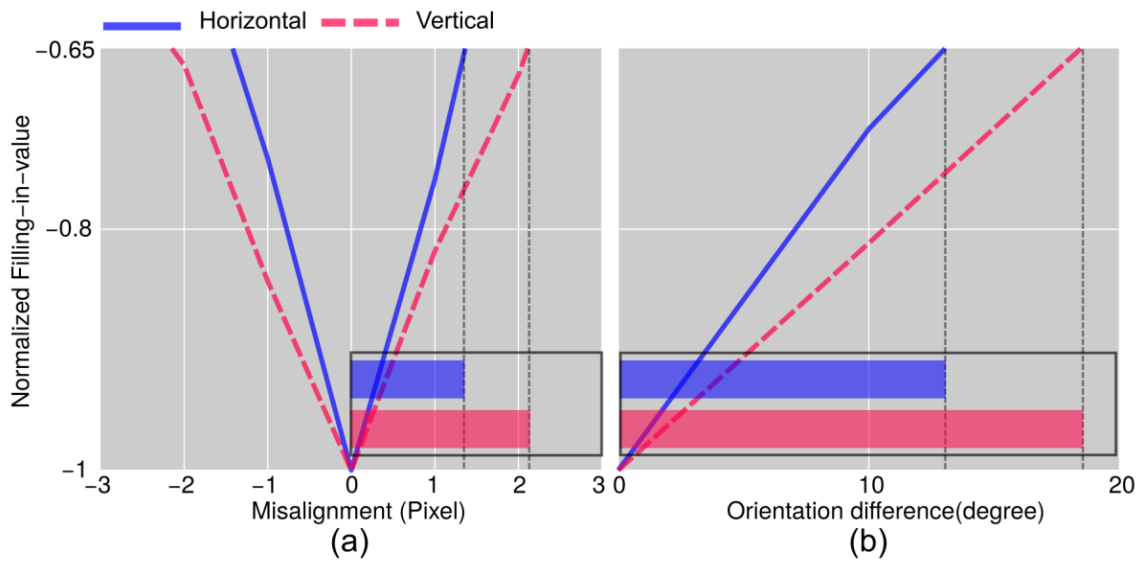

*Figure 7.  Anisotropy in tolerance of filing-in. Results presented in Fig. 5b and Fig. 6b has been redrawn after normalization in (a) and (b) respectively to compare with psychophysical results. Normalization has been done by dividing the respective curves by their respective magnitude of the maximum (more –ve) filling-in-value. The horizontal bars (in blue and red) shown in the inset (lower right of the (a) and lower left of the (b)) indicate the maximum difference in attribute (misalignment and orientation difference) for the horizontal and vertical configuration respectively beyond which filling-in is not perceivable.*

This process is similar to the one reported in[5] where the results of misalignment bar stimuli, identical to the results reported in[4] are presented along with the psychometric function. The threshold of perceptual completion was estimated from the value of the physical misalignment at 50% probability of the fitted psychometric function for perceptual completion at each orientation. These plots show that if the threshold estimation is conducted starting from the high probability (e.g. 100%) to 50% probability, the physical misalignment for completion will display an increasing trend.  This tendency is similar our results presented in Fig. 7 for misalignment as well as for orientation difference. The asymptotic shape of the plots near 100% probability[5] is not apparent in our results. This is possibly because of lesser resolution we have achieved in our simulation, where we have considered a $8\times 8$ pixels wide blind spot that provided 4 data points corresponding to four misalignments. For the same limited resolution, the results of collinear experiments[5] could not be predicted from our investigation. The collinear filling-in has been shown for the very narrow misalignment which is not possible in the current context. However, a model network with a better resolution could be able to exhibit similar results.

**Relation between natural image statistics and filling-in at the blind spot**

How anisotropy, then, arises from the response of the model network?  We have shown (Fig. 2b) that, in agreement with natural scene statistics, the distribution of the orientation preference of the learned receptive fields at V1 reflects the over-representation of neurons tuned towards horizontal orientation. This demonstrates that the model network could encode the anisotropies of natural scene statistics through learning. In a separate study[20] it has been suggested that the likelihood of filling-in of features (bars with different attributes) is guided by its likelihood of occurrence in the natural scene. Features that are more frequent tend to be more likely candidates for filling-in. In this perspective, we argue that the over-representation encoded by the learned receptive fields at V1 dominates the prediction at the blind spot that leads to filling-in of discontinuity. This happens because in the absence of the feed-forward connections (in the network representing blind spot region) top-down predictions biased by the learned internal model dominates.  Thus, the prevalence of horizontally oriented features (lines, bar etc.) in the learned internal model results in the superiority of horizontal features in filling-in. This is reflected as more negative 'filling-in-value' in all three horizontal cases (blue line) in Fig. 4b, Fig. 5b and Fig. 6b.

How vertical superiority arises in tolerance of filling-in? The nature of variation in filling-in-value, shown in Fig. 5b (or Fig. 7a) and Fig. 6b (or 7b), can be explained by taking into account the orientation tuning distribution of neurons shown in Fig. 2b. Inspection of Fig. 2b reveals that neurons tuned toward horizontal orientation have a higher population and sharper distribution. In comparison, neurons tuned toward vertical orientation have a relatively lower population and relatively broader distribution. The sharper distribution (and higher population) of neurons tuned toward horizontal orientation results in a more specific estimate for filling-in that would be less tolerance despite the fact that better filling-in will be observed for that orientation. On the other hand, broader distribution (and lower population) of neurons tuned toward vertical orientation results in higher tolerance and the lesser response results from the comparatively lower population. Therefore, in the case of horizontally oriented stimuli, the filling-in performance deteriorates at a faster rate with increasing difference in stimulus attributes compared to that of vertically oriented one.

These arguments can be readily put forward for explaining the anisotropy in tolerance of filling-in for disoriented bar stimuli (Fig. 6). For a given configuration (horizontal or vertical), the rotating segment of the stimuli makes varying angles with the fixed segment. Because of this, the filled-in section that resides inside the blind-spot will have to be aligned at varying angles either toward vertical or horizontal depending on the configuration. For every angle (0 to 90°), neurons having the similar orientation preference matching that of the filled-in section (in the blind spot) that connects the pair of bars will be activated for filling-in. For horizontal configuration, neurons having horizontal orientation preference as well as neurons having close to horizontal orientation preference are activated (depending on the stimuli in Fig. 6(b)). Because of the sharper distribution of neurons with orientation preference toward horizontal, a smaller orientation difference (with the horizontal) of the rotating bar will activate a certain population of neurons with similar orientation sensitivity but this population will be comparatively much smaller compared to the population that have been de-activated due to the increase in orientation difference. This will result in a larger decrease in response of the neurons, which is reflected as a faster decrease (lesser tolerance) in responses with increasing stimulus deviation from the horizontal orientation. Similar arguments can be given to explain the slower decrease (greater tolerance) in responses of neurons (because of broader distribution) in the case of vertical configuration.

In the case of misaligned bar investigation (Fig. 5), one bar is kept fixed and the other is shifted (either vertically or horizontally) to simulate varying amount of misalignment. Because of this, the filled-in section of the pair of bars (inside the blind-spot) will have to be aligned at varying angles either toward vertical or horizontal depending on the configuration. For every misalignment, neurons having orientation preference similar to that of the filled-in section become activated for filling-in. Therefore, as discussed before, the filling-in-value will be determined by the population of neurons tuned to a specific orientation and the nature of variation (with increasing misalignment) will be determined by the width of the distribution of neurons. This is reflected as better filling-in (more –ve filling-in-value) and faster deterioration in filling-in with increasing difference in attributes in case of the horizontal configuration shown in Fig. 5(b).

From the preceding discussions, it is evident that the predominance of horizontal contours in natural scene results in better filling-in operation in all three cases considered. This is reflected as more –ve filling-in-value as shown in Fig. 4, 5 and 6 (in blue). On the other hand, broader distribution of vertical contours results in a more tolerant response in filling-in operation with increasing difference in attributes. This is reflected in the curves (in red) with shallower gradient depicting the changing filling-in-value in Fig. 5 and 6.

Does the model HPC network predicts filling-in-values in accordance with statistics of natural images it was trained with? To validate these conclusions, we have repeated investigations with misaligned bar stimuli (Fig. 5) with a natural image and its 90° rotated version having vertical orientation superiority with asymmetric distribution of contours, which is shown in Fig. 8(a). The distribution of orientation content of the upper-left image is shown at the bottom of Fig. 8(a). We have evaluated the orientation at each pixel (upper left image in Fig. 8(a)) from the direction of the local gradient (of the grayscale image).

This was evaluated from the arc tangent of partial derivative (in $3\times3$ kernel) in the vertical direction divided by the value in the horizontal direction.

The distribution reveals the dominance of vertical contours and an asymmetric distribution around the dominant orientation (90 degrees) with a sharper rise (left side) and a slower fall (right side). Training with these two images produced an orientation preference of V1 neurons as shown in Fig. 8(c), where the neurons are equally sensitive to cardinal orientations and possessed similar distributions around cardinal orientations, which nearly preserved the asymmetries of the original image (Fig. 8(a)). This resulted in an equal filling-in response as shown by the superimposed curves (representing filling-in-values) in Fig. 8(d). Despite the fact that the distributions are similar, close inspection of Fig. 8 (c) reveals that the distributions, centered around cardinal angles, are asymmetric exhibiting a sharper rise at the left side and a comparatively slower fall at the right side. This implies that as long as the moving bar was aligned at $180-\theta$ ($90-\theta$) (Fig. 8(b)), the filling-in value altered at a faster rate with the angle and when it was aligned at $180+\theta$ ($90+\theta$), the filling-in value altered at a comparatively slower rate. This is reflected in the plot shown in Fig. 8(d) as faster rise on the left and a slower rise on the right side. From these results we conclude that the filling-in-value predicted by the model HPC network is in accordance with the statistics images used for training, where the absence of anisotropy in the dominance of the contours tuned to cardinal orientations results in equal filling-in response; and similar distribution of cardinal orientations results in similar gradient in the changing filling-in-value with increasing difference in the attributes.

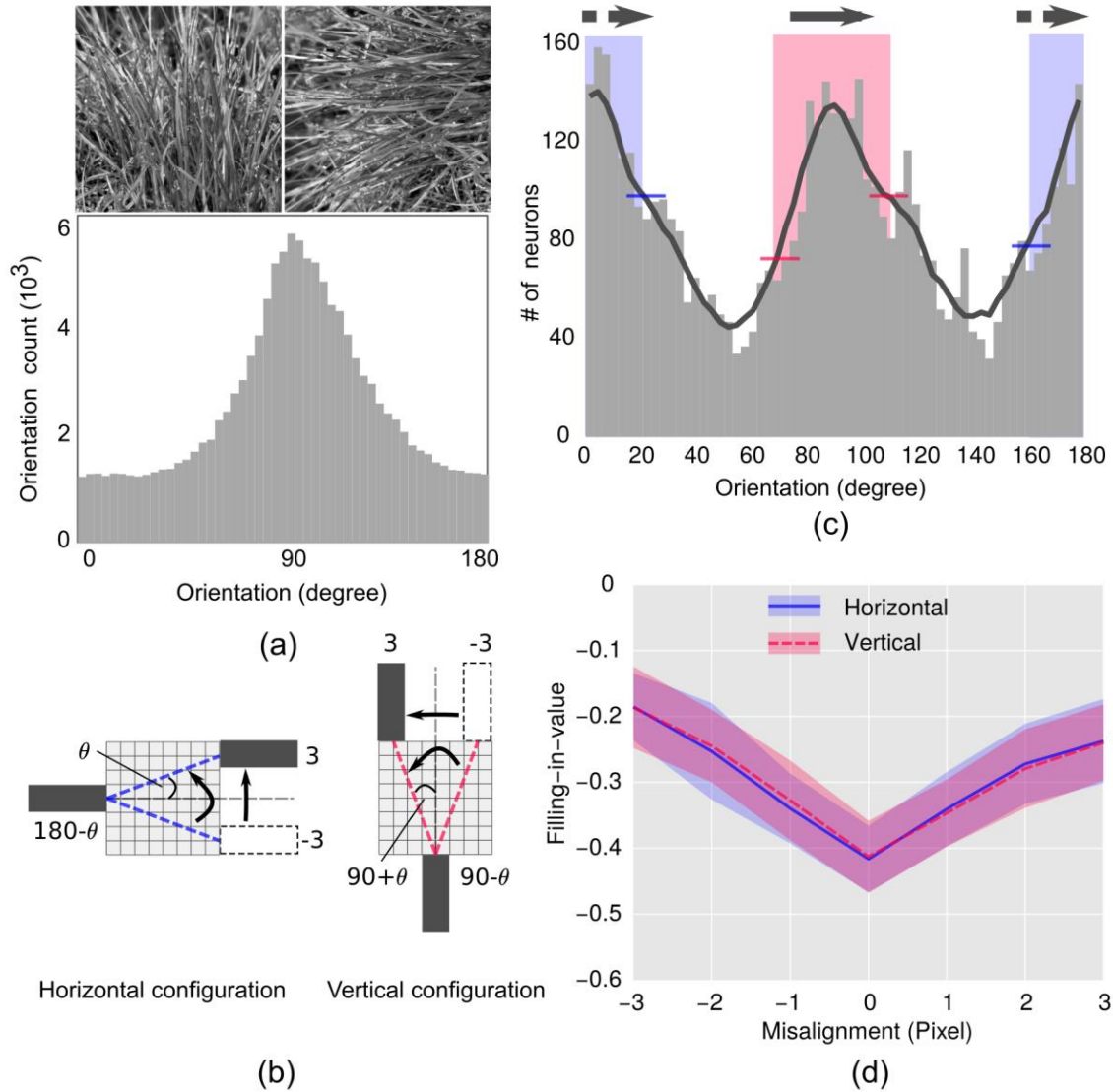

*Figure 8. Validation Investigation. (a) Natural images with asymmetric orientation distribution. The upper-left image mainly possesses contours with a bias towards vertical orientation. The histogram exhibiting this property is shown below. The upper-right image is 90 degrees rotated version of left one (histogram is not shown). (b) A detailed schematic of the misaligned bar study conducted in horizontal and vertical configuration. The moving bar was shifted by a maximum amount of 3 pixels on both sides of the mean (aligned) position. For the horizontal configuration it moved upward from the bottom and for the vertical case, it moved leftwards. The angular deviation of the filled-in portion (represented by dotted line inside the BS) can be evaluated from $\theta = \tan^{-1}$(position of the moving bar in pixels/8) (the size of BS area is $8 \times 8$ in pixels). (C) Orientation distribution of trained neurons at level 1. The continuous line (black) plot is the envelope of the histogram, which was obtained by convoluting the histogram, averaging over 7 bins. The shaded regions around horizontal (in blue) and vertical (in red) orientation indicate the population of neurons that is likely to be activated for filling-in when the moving bar is displaced by an amount $\pm 20$ degree ($\theta = \tan^{-1}(3/8) \sim 20$ degrees) around the mean position. The height difference between red lines (blue lines) across this smoothed plot is to indicate the neuronal density difference for the maximum misalignment (20 degrees) around the vertical orientation (horizontal orientation). The arrows above the shadowed regions indicate the direction of the moving bar. (d) Plots of 'filling-in-value' as a function of misalignment obtained from the response of the network. Convention for lines and the shades are as described in Fig. 4b.*

# Discussions

Our study suggests that natural scene statistics plays a significant role in determining the anisotropy in perceptual filling-in including the anisotropy in tolerance of perceptual filling-in at the blind spot. Over-representation of horizontal contours in natural scene biases the orientation preference of neurons in V1 and that is possibly responsible for the emergence of anisotropy, which is reflected as a horizontal preference in perceptual filling-in operation. The width of the distribution of orientation preference, on the other hand, determines the anisotropy in tolerance of filling-in, where the broader distribution of vertical contours in natural scene possibly contributes to the greater stability towards vertical orientation in perceptual filling-in operation.

These results demonstrate that there is a link between the orientation anisotropy in the contours in the natural environment, orientation preference of neurons in V1 and orientation bias in the perceptual filling-in at the blind spot. Our result supports the general speculation[6,8,10–12] that orientation anisotropy in natural scene plays a significant role in determining the anisotropy in the cortex as well as the anisotropy in perceptual orientation preference.

Firstly, we show that the model HPC network, which mimics the prediction-correction computational paradigm of the cortex, is capable of building an internal model of the outside environment by learning the statistics of natural scenes it is exposed to. This is reflected by the fact that the orientation preference, as well as the distribution of orientation preference of model neurons in V1, is very similar to the predominance of horizontal contours and their distribution in the natural environment. The plausibility of this paradigm can be established with the help of several previous findings. In a recent survey[16], in the physiological domain, involving cells in the cat's striate cortex indicate the preferential bias of cells towards horizontal orientation. Imaging studies also revealed[17–19] the preference of higher percentage of the area of the exposed visual cortex towards horizontal orientation compared to vertical. Innate specification along with prolonged exposure to an anisotropic environment during development is believed to be responsible for the emergence of overrepresentation of horizontal orientation preference of these neurons. In the psychophysical domain, correspondence between the horizontal bias in human visual processing and the anisotropy in the natural scene has been reported in[8,9]. A detailed survey in this work also shows the prevalence of horizontal contours in a typical natural scene compared to vertical contours. In a recent study, it has been demonstrated that visual orientation perception reflects the knowledge of environmental statistics[6]. In this work, the estimated internal model of human observers was found to match the orientation distribution measured in photographs of environment though the difference between horizontal and vertical was not addressed.

Secondly, our investigations reveal that the anisotropy in orientation preference (horizontal) of V1 neurons results in the similar anisotropy in the filling-in performance and the distribution (sharper or broader) of cardinal neurons results in the anisotropy of tolerance in filling-in performance. What is the biological plausibility of such a scheme? In an imaging study[13] it has been shown that in V1 the distribution of inputs to the cardinal neurons is narrower compared to those of oblique neurons. When exposed to selective perturbation induced by adaptation (oriented away from the neuron's preferred orientation), cardinal neurons exhibited greater stability compared to the neurons tuned to oblique orientation. This is attributed to the fact that because of the narrower distribution of local inputs to the cardinal neurons, an adaptive stimulus would stimulate a fewer number of neurons in the vicinity compared to that of the neurons tuned to oblique orientation. This demonstrates that the width of the distribution (of neurons) plays a significant role in determining the responses when stimulated away from the preferred orientation. From a different perspective it indicates that for neurons having narrower distribution, a much greater change in response will be observed with increasing deviation of the stimulus orientation from the neuron's preferred orientation. This implies greater sensitivity and therefore, lesser stability in the present context. Comparatively, neurons having broader distribution will be less sensitive (more stable). This is similar to the findings of our observation. Evidence in favour of larger neural population preferring horizontal orientation (compared to vertical) have also been found in several physiological studies[16–18], as discussed earlier.

In studies on filling-in completions at the blind spot[4,5], it was speculated that there might be different anisotropic process responsible for different kinds of anisotropy observed in different (misalignment, disorientation, and luminance difference) filling-in investigations e.g., it was speculated that the anisotropy in misalignment experiment might have arose from the anisotropy in vernier acuity. Here in this study, we have proposed a possible alternative explanation in terms of a unified principle based on the role natural image statistics. We have demonstrated this in filling-in investigations involving misaligned and disoriented bar stimuli. Results of our studies also suggest that the anisotropy in vernier acuity might have its origin in the statistics of natural scenes. Evidence in support of these suggestions can be found in[26], where it was argued that the vernier misalignment can be discussed on the premise that the average orientation of a misaligned pair of abutting lines differs from that of the aligned lines. Vernier acuity preferring horizontal directions over the vertical including the cardinal over the oblique has been demonstrated in this work.

We speculate that the horizontal superiority[4] in the tolerance of luminance difference could be discussed in terms of statistics of the natural scene. Luminance is a surface property, and, therefore, for proper inference, the cortex should be capable of encoding 3D surface information efficiently. In a recent study[27] it has been shown that disparity neurons are capable of encoding statistics of the natural scene. Studies[28] also show that the pair-wise functional connectivity between the disparity tuned neurons in V1 matches the anisotropic distribution of correlation between disparity signals in natural scene. Though, these studies mainly concentrated on the cardinal vs non-cardinal aspect of the anisotropy, a close inspection of the plots indicate a broader distribution of the horizontal features. This broader distribution in disparity signal (or pair-wise connectivity) could be linked to the horizontal superiority in the tolerance of luminance difference. Some supportive evidence can be found in a recent work[29] showing that relative luminance and binocular disparity preferences are correlated in accordance with the trends of natural scene statistics. These studies suggest a possible link between the anisotropy in the disparity signal and the relative luminance. In a future work, incorporation of surface representation in the internal model in the HPC framework might explain the anisotropy in luminance difference.

In this work, we have investigated the origin of anisotropy in perceptual filling-in in a simple standard linear Hierarchical Predictive Coding network. Because of this, our findings could only explain the possible reasons responsible for the emergence of anisotropy in filling-in reported by human participants, but a quantitative comparison with psychophysical results is not straight forward. In the present context, however, what matters is that given the statistical information of the input stimuli derived from natural images, the network was able to predict the anisotropy in perceptual filling-in at the blind spot. The findings, in this work, offer new insights into the role of natural scene statistics and suggest what is possibly the first systematic bridge linking anisotropy in three levels: natural environment, visual cortex, and perceptual filling-in at the blind spot.

# Methods

## *Standard hierarchic predictive coding (HPC):*

In this paradigm, the visual system is considered to be an active predictor-corrector system implemented in a hierarchical neural architecture where perception is accomplished via the interaction of top-down prediction and bottom-up correction[21,30]. Instead of passively responding to the input signal, higher-level cortical activities (predictions) are conveyed to lower levels via top-down connections and in response, lower levels convey residual errors via bottom-up connections (see Fig. 9a). It is further assumed that prediction by the higher cortical levels is mainly governed by the regularities learned via the exposure to the natural scene during development.

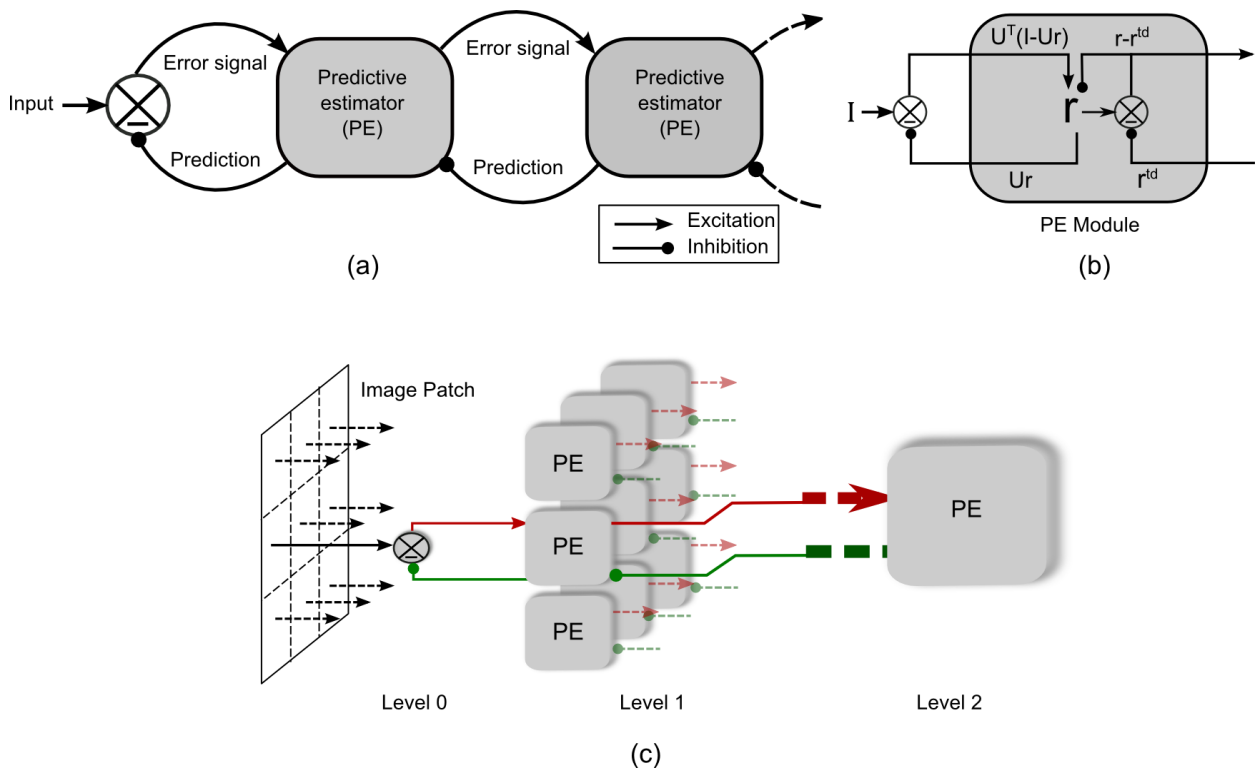

*Figure 9. Hierarchical predictive coding (HPC)[20] (a) General mechanism of HPC. (b) General computational architecture of a predictive estimator (PE) module. (c) A three level HPC model network; where level 2 module sends a feedback signal to all 9 level 1 modules and in response, receives back the error signal from all of them.*

The visual system learns the model of the outer world through its parameters related to statistical regularities $\mathbf{U}$. The prediction $\mathbf{Ur}$ is generated from the activity of the neurons coding the internal representations or estimate $\mathbf{r}$ which is the actual cause of input sensory signal $\mathbf{I}$. Given that the vision is a stochastic phenomenon, the goal of the visual system is, thus, to maximize the posterior probability distribution $P(\mathbf{r}, \mathbf{U} \mid \mathbf{I})$. According to Bayesian theorem, this is roughly equal to the product of likelihood $P(\mathbf{I} \mid \mathbf{r}, \mathbf{U})$, which is a distribution of stochastic error between prediction and sensory input, and the prior distributions $P(\mathbf{r})$ and $P(\mathbf{U})$. Assuming Gaussian type stochastic error, with variance $\sigma^2$, the posterior distribution can be written as -

$$P(\mathbf{r}, \mathbf{U} \mid \mathbf{I}) = \frac{1}{Z} \exp(-\frac{|\mathbf{I} - \mathbf{U}\mathbf{r}|^2}{\sigma^2}) P(\mathbf{r}) P(\mathbf{U}) \tag{1}$$

Where, $Z$, is a normalization constant. Maximizing this equation is equivalent to minimizing the negative logarithm of it, which is called cost function in the MDL terminology and can be written as -

$$E_1 = \frac{1}{\sigma^2}(\mathbf{I} - \mathbf{U}\mathbf{r})^T(\mathbf{I} - \mathbf{U}\mathbf{r}) + g(\mathbf{r}) + h(\mathbf{U}) \tag{2}$$

Where subscript $T$ indicates the transpose of the vector or matrix. In addition $g(\mathbf{r})$, $h(\mathbf{U})$ are the negative logarithm of $P(\mathbf{r})$ and $P(\mathbf{U})$, respectively.

The cost function of an inference system with 3 level of hierarchy, in which the higher (3rd) level makes inference (or prediction) $\mathbf{r}^{td}$ to the immediate level representation $\mathbf{r}$ with error variance $\sigma_{td}^2$, can be written as (for details see [21,30])

$$E_2 = \frac{1}{\sigma^2}(\mathbf{I} - \mathbf{U}\mathbf{r})^T(\mathbf{I} - \mathbf{U}\mathbf{r}) + \frac{1}{\sigma_{td}^2}(\mathbf{r} - \mathbf{r}^{td})^T(\mathbf{r} - \mathbf{r}^{td}) + g(\mathbf{r}) + h(\mathbf{U}) \tag{3}$$

This equation serves as a guiding principle for the standard Hierarchical Predictive Coding (see Fig. 9b), which assumes that the predictive estimator (PE) modules at each visual processing level send the prediction signal $\mathbf{U}\mathbf{r}$ to its immediate lower processing level via feedback connection. On the other hand, the lower levels send back the error signal $(\mathbf{I} - \mathbf{U}\mathbf{r})$ via feed-forward connection. The error signal is then utilized to correct the current estimate $\mathbf{r}$, which is coded by PE neurons, of the sensory driven input.

The dynamics and the learning rule, thus, result from minimizing the cost function (using gradient decent method), with respect to $\mathbf{r}$ and $\mathbf{U}$ respectively-

$$\frac{d\mathbf{r}}{dt} = -\frac{k_1}{2}\frac{\partial E_2}{\partial \mathbf{r}} = \frac{k_1}{\sigma^2}\mathbf{U}^T(\mathbf{I} - \mathbf{U}\mathbf{r}) + \frac{k_1}{\sigma_{td}^2}(\mathbf{r}^{td} - \mathbf{r}) - \frac{k_1}{2}g'(\mathbf{r}) \tag{4}$$

$$\frac{d\mathbf{U}}{dt} = -\frac{k_2}{2}\frac{\partial E_2}{\partial \mathbf{U}} = \frac{k_2}{\sigma^2}(\mathbf{I} - \mathbf{U}\mathbf{r})\mathbf{r}^T - \frac{k_2}{2}h'(\mathbf{U}) \tag{5}$$

Kurtosis prior probability $P(r_i) = \exp(-\alpha \ln(1 + r_i^2))$ on response $r_i$ has been considered in this study to accommodate sparse coding[31], which provides $g'(r_i) = 2\alpha r_i/(1 + r_i^2)$. Additionally, considering prior $P(\mathbf{U})$ as a Gaussian, provides $h'(\mathbf{U}) = 2\lambda \mathbf{U}$. Here $\alpha$ and $\lambda$ are variance related parameters.

An optimum estimate at a visual processing level is determined by the error signal from lower area (first term in the equation (4)) as well as error signal corresponding to a higher level (second term in equation (4)) that carry the contextual information since the higher area codes larger visual patch. This multilevel optimum-estimate for prediction is considered as an internal representation of the sensory input. The internal representation fabricated from the prediction $\mathbf{U}\mathbf{r}$ is assumed to represent 'perceptual experience' in this study.

*Network:*

A three-level network has been used in this study (Fig. 9c). Level 0, level 1 and level 2 are equivalent to the LGN, V1, and V2. Level 0 pre-process (low pass filtering) the stimuli in line with LGN function. Each module at level 1 sends prediction signal to level 0, by feedback connection and in response receives the error signal by the feed-forward connection. Likewise, each module at level 2 sends the prediction signal to all 9 modules at level 1, and get back the error signal by a feed-forward connection from all of them. The modules at level 1 consist of 130 feed-forward, 130 PE neurons, and 144 feedback neurons. The level 2 module contains 256 feed-forward neurons, 256 PE neurons, and 1170 feedback neurons.

Training:

For obtaining statistically significant results, we performed 40 training cycles. In each training cycle, the network receives a thousand batches of 100, variance normalized, pre-processed[31] $30 \times 30$-pixel image patches as inputs. Each level 1 module receives signal corresponding to $12 \times 12$-pixel image patches which were overlapped by 3 pixels[20]. The network was allowed to achieve the optimum-estimate (equation (4)) for each batch and then the average of the optimum-estimate was used to update the weighting profile of neurons (equation (5)), initially assigned to random values. To prevent the weighting profile from growing boundlessly, the gain of the weighting profile of each neuron were adapted such that it maintains the equal variances on the response. Parameters used in this study are same as considered in the previous study[20].

*Blind spot implementation:*

First, the model network was trained without considering the blind spot, and thereafter, the blind spot was created in the trained network by removing the feed-forward connection from level 0 to level 1 ($8 \times 8$ pixel wide in the middle of BS module). This process is in agreement with the actual physiological findings, where the neurons contributing to the filling-in process (at the blind spot) are found to be of binocular type and therefore, receive inputs from both the eye. Thus, in spite of the absence of any input from one eye (the blind spot eye), the neurons could develop their weighting profiles. For a detailed discussion see the previous study[20]

# Author contributions

R.R. and S.S. conceived the study. R.R. performed the model simulations. R.R. and S.S. wrote the manuscript.

# Competing financial interests

We have no competing financial interest to declare.